\documentclass[12pt,a4paper]{article}
\usepackage{amssymb}

\usepackage{graphicx}
\usepackage{amsmath}
\usepackage{makeidx}
\usepackage{graphicx}
\usepackage{graphicx}
\usepackage{indentfirst}

\newcounter{resultnum}[section]

\newcounter{conclusionnum}[section]

\newcounter{conditionnum}[section]

\newcounter{conjecturenum}[section]

\newcounter{examplenum}[section]

\newcounter{exercisenum}[section]

\newcounter{lemmanum}[section]

\newcounter{notationnum}[section]

\newcounter{theoremnum}[section]

\newcounter{definitionnum}[section]

\newcounter{corollarynum}[section]

\newcounter{remarknum}[section]

\newcounter{propositionnum}[section]

\newcounter{acknowledgementnum}[section]

\newcounter{algorithmnum}[section]

\newcounter{axiomnum}[section]

\newcounter{casenum}[section]

\newcounter{claimnum}[section]

\newcounter{summarynum}[section]

\newcounter{problemnum}[section]

\begin{document}

\title{Nonholonomic Ricci Flows and \\  Running Cosmological Constant: \\
3D Taub--NUT Metrics}

\author{ Sergiu I. Vacaru\thanks{sergiu$\_$vacaru@yahoo.com,
 svacaru@brocku.ca, svacaru@fields.utoronto.ca } \\
{\small \it The Fields Institute for Research in Mathematical Science}\\
{\small \it 222 College Street, 2d Floor, Toronto, Ontario M5T 3J1, Canada}
\and
Mihai Visinescu\thanks{mvisin@theory.nipne.ro}\\
{\small \it  Department of Theoretical Physics,}\\
{\small \it National Institute for Physics and Nuclear Engineering} \\
{\small \it P.O. Box M.G.-6, Magurele, Bucharest, Romania}}

\date{October 11, 2007}
\maketitle

\begin{abstract}
The common assertion that the Ricci flows of Einstein spaces with
cosmological constant can be modelled by certain classes of nonholonomic
frame, metric and linear connection deformations resulting in
nonhomogeneous Einstein spaces is examined in the light of the role
played by topological three dimensional (3D) Taub--NUT--AdS/dS spacetimes.

\vskip0.3cm



\end{abstract}

\newpage 

\tableofcontents

\section{ Introduction: Geometric Preliminaries}

There is a growing interest in the geometry of Ricci flows \cite%
{ham1,ham2,aubin,cao,chen} and its applications in high energy physics and
cosmology \cite{per,bakas,geg,dm,wg,hw}. This prompts one to study
systematically
various examples of exactly solvable models. This paper is a continuation of
 \cite{rf4de}\footnote{%
See there detailed motivations and outlines of the anholonomic frame
method.}%
, extending our previous works \cite{crvis} and \cite{vrf}, oriented to a
research on three dimensional (3D) Ricci flows with nonholonomic
deformations of Thurston's geometries or topological Taub-NUT--AdS/dS
(anti--de Sitter/de Sitter) configurations \cite{thurs,ms}. The 3D case is
the simplest one when exact solutions for nonholonomic Ricci flow equations
can be constructed in explicit form and related to physically important
exact solutions of the Einstein equations. Such 3D metrics are characterized
by corresponding nonholonomic symmetries and, in general, possess nontrivial
torsion induced by nonholonomic deformations and/or from string gravity.

In this paper, we shall study 3D Ricci flows of off--diagonal metrics
trivially embedded into a 4D spacetime,
\begin{equation}
\mathbf{g}=g_{\underline{\alpha }\underline{\beta }}(u)du^{\underline{\alpha
}}\otimes du^{\underline{\beta }},  \label{metr}
\end{equation}%
where
\begin{equation*}
g_{\underline{\alpha }\underline{\beta }}=\left[
\begin{array}{cc}
g_{ij}+N_{i}^{a}N_{j}^{b}h_{ab} & N_{j}^{e}h_{ae} \\
N_{i}^{e}h_{be} & h_{ab}%
\end{array}%
\right] ,
\end{equation*}%
with the indices of type $\underline{\alpha },\underline{\beta }%
=(i,a),(j,b)...$ running the values $i,j,...=1,2$ and $a,b,...=3,4$ (we
shall omit underlying of indices for the components with respect to
coordinate basis if that will not result in ambiguities) and local
coordinates labeled in the form $u=(x,y)=\{u^{\alpha }=(x^{i},y^{a})\}.$ In
order to preserve a unique system of denotations together with Ref. \cite%
{rf4de}, we shall consider parametrizations when $g_{11}=\epsilon =\pm 1$
and $N_{1}^{a}=0,$ $g_{22}=g_{22}(x^{2}),$ $%
h_{ae}=diag[h_{3}^{a}(x^{2},y^{b}),h_{4}^{a}(x^{2},y^{b})]$ and $%
N_{2}^{a}=N_{2}^{a}(x^{2},y^{b}),$ i.e. when the coordinate $x^{1}$ will not
be contained into further ansatz for metric and connections. This formal
convention will allow to apply directly a number of results considered in 4D
gravity and related Ricci flow solutions (we shall omit proofs and details
in such cases and we shall refer the reader to the corresponding works where
similar cases were analyzed for 4D or 5D constructions).

We write the normalized Ricci flow equations \cite%
{ham1,ham2,aubin,cao,per,bakas} in the form
\begin{equation}
\frac{\partial }{\partial \tau }g_{\underline{\alpha }\underline{\beta }%
}=-2R_{\underline{\alpha }\underline{\beta }}+\frac{2r}{5}g_{\underline{%
\alpha }\underline{\beta }},  \label{feq}
\end{equation}%
where $R_{\underline{\alpha }\underline{\beta }}$ is the Ricci tensor of a
metric $g_{\underline{\alpha }\underline{\beta }}$ and corresponding Levi
Civita connection (by definition this connection is torsionless and metric
compatible) and the normalizing factor $r=\int RdV/dV$ is introduced in
order to preserve the volume $V.$ It should be emphasized that in our works
\cite{rf4de,vrf} the running parameter $\tau$ is treated as a space--time
coordinate
(usually being time like or extra dimension one), which is naturally in order
to relate the constructions with physical models on pseudo--Riemannian spaces.
In this case, it is a more difficult task to construct exact solutions.
For flows on Riemannian manifolds with $\tau$ considered as an "external"
parameter labeling families of metrics, the method
 of generating nonholonomic exact solutions simplifies substantially.

We can represent the metric (\ref{metr}) in effectively diagonalized $%
(1+1+2) $--distingu\-ish\-ed form
\begin{equation}
\mathbf{g}=g_{\alpha }(u)\ \mathbf{c}^{\alpha }\otimes \mathbf{c}^{\alpha
}=\epsilon \ b^{1}\otimes b^{1}+g_{2}(x^{2})\ b^{2}\otimes b^{2}+h_{a}(u)\
b^{a}\otimes b^{a},  \label{m2a}
\end{equation}%
with respect the basis
\begin{equation}
\mathbf{c}^{\alpha }=(b^{i}=dx^{i},b^{a}=dy^{a}+N_{2}^{a}(u)dx^{2})
\label{ddif}
\end{equation}%
being dual to the local basis
\begin{equation}
\mathbf{e}_{\alpha }=(e_{1}=\frac{\partial }{\partial x^{1}},e_{2}=\frac{%
\partial }{\partial x^{2}}-N_{2}^{b}(u)\frac{\partial }{\partial y^{b}}%
,e_{b}=\frac{\partial }{\partial y^{b}}),  \label{dder}
\end{equation}%
where $dx^{1}$ and $\partial _{1}=\partial /\partial x^{1}$ are not
considered in the case of 3D configurations. Such metric parametrizations
and frame transforms have been introduced in the geometry of nonholonomic
manifolds with associated N--connection structure defined by the set $%
\mathbf{N}=\{N_{k}^{b}\}$ stating a nonholonomic preferred local frame on a
3D manifold $\mathbf{V}.$ We shall examine Ricci flows of 3D metrics
parametrized by ansatz of type (\ref{m2a}) when
\begin{equation*}
g_{2}=g_{2}(x^{2}),h_{a}=h_{a}(x^{2},v),N_{2}^{3}=
w_{2}(x^{2},v),N_{2}^{4}=n_{2}(x^{2},v),
\end{equation*}%
for $y^{3}=v$ being the so--called ''anisotropic'' coordinate.

In order to consider flows of metrics related both to the Einstein and
string gravity (in the last case there is a nontrivial antisymmetric torsion
field), it is convenient to work with the so--called canonical distinguished
connection (in brief, d--connection) $\widehat{\mathbf{D}}=\{\widehat{%
\mathbf{\Gamma }}_{\ \alpha \beta }^{\gamma }\}$ which is metric compatible
but with nontrivial torsion (see formulas (54) and related discussions for
formulas (53), (56)-- (60) in Appendix to \cite{rf4de}). Imposing certain
restrictions on the coefficients $N_{k}^{b},$ we can satisfy the conditions
that the coefficients of the canonical d--connection of the Levi-Civita $%
\bigtriangledown =\{\ _{\shortmid }\Gamma _{\beta \gamma }^{\alpha }\}$ are
defined by the same nontrivial values $\widehat{\mathbf{\Gamma }}_{\ \alpha
\beta }^{\gamma }=\ _{\shortmid }\Gamma _{\ \alpha \beta }^{\gamma }$ with
respect to N--adapted basis (\ref{dder}) and (\ref{ddif}).

The Ricci flow equations (\ref{feq}) can be written for the Ricci tensor of
the canonical d--connection, $R_{\alpha \beta }=[R_{22},R_{a2},S_{ab}],$ and
metric (\ref{m2a}), as it was considered in Refs. \cite{vrf,rf4de},
\begin{eqnarray}
\frac{\partial }{\partial \tau }g_{2} &=&-2R_{22}+2\lambda g_{2}-h_{cd}\frac{%
\partial }{\partial \tau }(N_{2}^{c}N_{2}^{d}),  \label{1eq} \\
\frac{\partial }{\partial \tau }h_{ab} &=&-2S_{ab}+2\lambda h_{ab},\
\label{2eq} \\
R_{\alpha \beta } &=&0\mbox{ and }g_{\alpha \beta }=0\mbox{ for }\alpha \neq
\beta ,\text{ }  \notag
\end{eqnarray}%
where $\lambda =r/5,$ $y^{3}=v$ and $\tau $ can be, for instance, the time
like  coordinate, $\tau =t,$ or any parameter or extra dimension
coordinate. The equations (\ref{1eq}) and (\ref{2eq}) are just the
nonholonomic transform of the Ricci equations (\ref{feq}) if there are
imposed such constraints that $\widehat{\mathbf{\Gamma }}_{\ \alpha \beta
}^{\gamma }=\ _{\shortmid }\Gamma _{\ \alpha \beta }^{\gamma }$. The aim of
this work is to show how the anholonomic frame method developed in \cite%
{vrf,rf4de} (for Ricci flows) and in \cite{vhep2,vs,vesnc,vt,vp} (for
off--diagonal exact solutions) can be used for constructing exact solutions
of the system of Ricci flow equations (\ref{1eq}) and (\ref{2eq}) describing
nonholnomic deformations of 3D Taub-NUT solutions.

The structure of the paper is as follows: In section 2 we apply the
anholonomic frame method in the geometry of Ricci flows of 3D off--diagonal
metrics. We construct a general class of integral varieties of Ricci flow
equations and define the constrains for the Levi-Civita configurations.
Section 3 is devoted to explicit solution of nonholonomic 3D Ricci flows and
concentrate on two examples of constructing solutions starting with primary
Taub--NUT--dS metrics. We consider details on derivation of solutions and
analyze flows on holonomic and anholonomic time like coordinate. The last
section is devoted to conclusions and discussion.

\section{Geometry of 3D Off--Diagonal Ricci Flows}

In this section, we show how following the anholonomic frame method the 3D
Ricci flow equations can be integrated in a general form.

\subsection{Ricci flow equations for off--diagonal metric ansatz}

The nontrivial components of the Ricci tensor $R_{\alpha \beta }$ (see
details of a similar calculus in Ref. \cite{vesnc}) are
\begin{eqnarray}
S_{\ 3}^{3} &=&S_{\ 4}^{4}=\frac{1}{2h_{3}h_{4}}\left[ -h_{4}^{\ast \ast }+%
\frac{\left( h_{4}^{\ast }\right) ^{2}}{2h_{4}}+\frac{h_{4}^{\ast
}h_{3}^{\ast }}{2h_{4}}\right] ,  \label{rtv} \\
R_{32} &=&-\frac{1}{2h_{5}}\left( w_{2}\beta +\alpha _{2}\right) ,
\label{3eq} \\
R_{42} &=&-\frac{h_{5}}{2h_{4}}\left( n_{2}^{\ast \ast }+\gamma n_{2}^{\ast
}\right)  \label{4eq}
\end{eqnarray}%
where
\begin{eqnarray}
\alpha _{2} &=&\partial _{2}h_{4}^{\ast }-h_{4}^{\ast }\partial _{2}\ln
\sqrt{\left| h_{3}h_{4}\right| },\ \beta =h_{4}^{\ast \ast }-h_{4}^{\ast
}\left( \ln \sqrt{\left| h_{3}h_{4}\right| }\right) ^{\ast },  \label{aux1}
\\
\gamma &=&3h_{4}^{\ast }/2h_{4}-h_{3}^{\ast }/h_{3},\mbox{ for }h_{3}^{\ast
}\neq 0,h_{4}^{\ast }\neq 0,  \notag
\end{eqnarray}%
defined by $h_{3}$ and $h_{4}$ as solutions of equations (\ref{2eq}). In the
above presented formulas, it was convenient to write the partial derivative
on the so--called "anisotropic" coordinate $v$ in the form $a^{\ast
}=\partial a/\partial v.$

We consider a general method of constructing solutions of the Ricci flows
equations related to so--called Einstein spaces with nonhomogeneously
polarized cosmological constant, when
\begin{eqnarray}
S_{\ b}^{a} &=&\lambda _{\lbrack v]}(x^{2},v)\ \delta _{\ b}^{a},  \notag \\
R_{\alpha \beta } &=&0\mbox{ and }g_{\alpha \beta }=0\mbox{ for }\alpha \neq
\beta ,\text{ }  \notag
\end{eqnarray}%
with $\lambda _{\lbrack v]}$ induced by certain string gravity ansatz, or
matter field contributions, see \cite{vrf,rf4de}.

The nonholonomic Ricci flows equations (\ref{1eq}) and (\ref{2eq}) for the
Einstein spaces with nonhomogeneous cosmological constant defined by ansatz
of type (\ref{m2a}) transform into the following system of partial
differential equations consisting from two subsets of equations: The first
subset of equations consists from those generated by the 3D Einstein
equations for the off--diagonal metric,
\begin{eqnarray}
-h_{4}^{\ast \ast }+\frac{\left( h_{4}^{\ast }\right) ^{2}}{2h_{4}}+\frac{%
h_{4}^{\ast }h_{3}^{\ast }}{2h_{4}} &=&2h_{3}h_{4}\ \lambda _{\lbrack
v]}(x^{2},v),  \label{ee2} \\
w_{2}\beta +\alpha _{2} &=&0,  \label{ee3} \\
n_{2}^{\ast \ast }+\gamma n_{2}^{\ast } &=&0.  \label{ee4}
\end{eqnarray}%
The second subset of equations is formed just by those describing flows of
the diagonal, $g_{ij}=diag[\epsilon ,g_{2}]$ and $h_{ab}=diag[h_{3},h_{4}],$
and off diagonal, $w_{2}$ and $n_{2},$ metric coefficients,
\begin{eqnarray}
\frac{\partial }{\partial \tau }g_{2} &=&-h_{3}\frac{\partial }{\partial
\tau }\left( w_{2}w_{2}\right) -h_{4}\frac{\partial }{\partial \tau }\left(
n_{2}n_{2}\right) ,  \label{rfe1} \\
\frac{\partial }{\partial \tau }h_{a} &=&2\lambda _{\lbrack v]}(x^{2},v)\
h_{a}.  \label{rfe2}
\end{eqnarray}%
The aim of the next section is to show how we can integrate the equations (%
\ref{ee2})--(\ref{rfe2}) in a quite general form.

\subsection{Integral varieties for 3D Ricci flow equations}

The equation (\ref{ee2}) relates two nontrivial v--coefficients of the
metric coefficients $h_{3}(x^{2},v)$ and $h_{4}(x^{2},v)$ depending on three
coordinates but with partial derivatives only on the third (anisotropic)
coordinate. As a matter of principle, we can fix $h_{3}$ (or, inversely, $%
h_{4}$) to describe any physically interesting situation being, for
instance, a solution of the 3D solitonic, or pp--wave equation,  and than
we can try to define $h_{4}$ (inversely, $h_{3}$) in order to get a solution
of (\ref{ee2}). Here we note that it is possible to solve such equations for
any $\lambda _{\lbrack v]}(x^{2},v),$ in general form, if $h_{4}^{\ast }\neq
0$ (for $h_{4}^{\ast }=0,$ there are nontrivial solutions only if $\lambda
_{\lbrack v]}=0).$ Introducing the function
\begin{equation}
\phi (x^{2},v)=\ln \left| h_{4}^{\ast }/\sqrt{|h_{3}h_{4}|}\right| ,
\label{afa1}
\end{equation}%
we write that equation in the form%
\begin{equation}
\left( \sqrt{|h_{3}h_{4}|}\right) ^{-1}\left( e^{\phi }\right) ^{\ast
}=-2\lambda _{\lbrack v]}.  \label{afa2}
\end{equation}%
Using (\ref{afa1}), we express $\sqrt{|h_{3}h_{4}|}$ as a function of $\phi $
and $h_{4}^{\ast }$ and obtain%
\begin{equation}
|h_{4}^{\ast }|=-(e^{\phi })^{\ast }/4\lambda _{\lbrack v]}  \label{fa1}
\end{equation}%
which can be integrated in general form,
\begin{equation}
h_{4}=h_{4[0]}(x^{i})-\frac{1}{4}\int dv\ \frac{\left[ e^{2\phi (x^{2},v)}%
\right] ^{\ast }}{\lambda _{\lbrack v]}(x^{2},v)},  \label{solh4}
\end{equation}%
where $h_{4[0]}(x^{2})$ is the integration function. Having defined $h_{4}$
and using again (\ref{afa1}), we can express $h_{3}$ via $h_{4}$ and $\phi ,$
\begin{equation}
|h_{3}|=4e^{-2\phi (x^{2},v)}\left[ \left( \sqrt{|h_{4}|}\right) ^{\ast }%
\right] ^{2}.  \label{solh5}
\end{equation}%
The conclusion is that prescribing any two functions $\phi (x^{2},v)$ and $%
\lambda _{\lbrack v]}(x^{2},$ $v)$ we can always find the corresponding
metric coefficients $h_{3}$ and $h_{4}$ solving (\ref{ee2}). Following (\ref%
{solh5}), it is convenient to represent such solutions in the form%
\begin{eqnarray*}
h_{4} &=&\epsilon _{4}\left[ b(x^{2},v)-b_{0}(x^{2})\right] ^{2} \\
h_{3} &=&4\epsilon _{3}e^{-2\phi (x^{2},v)}\left[ b^{\ast }(x^{2},v)\right]
^{2}
\end{eqnarray*}%
where $\epsilon _{a}=\pm 1$ depending on fixed signature, $b_{0}(x^{2})$ and
$\phi (x^{2},v)$ can be arbitrary functions and $b(x^{2},v)$ is any function
when $b^{\ast }$ is related to $\phi $ and $\lambda _{\lbrack v]}$ as stated
by the formula (\ref{fa1}). Finally, we note that if $\lambda _{\lbrack
v]}=0,$ we can relate $h_{3}$ and $h_{4}$ solving (\ref{afa2}) as $\left(
e^{\phi }\right) ^{\ast }=0.$

For any couples $h_{3}$ and $h_{4}$ related by (\ref{ee2}), we can compute
the values $\alpha _{2},\beta $ and $\gamma $ (\ref{aux1}). This allows us
to define the off--diagonal metric (N--connection) coefficients $w_{2}$
solving (\ref{ee3}) as algebraic equations,%
\begin{equation}
w_{2}=-\alpha _{2}/\beta =-\partial _{2}\phi /\phi ^{\ast }.  \label{see3}
\end{equation}%
We emphasize, that for the vacuum Einstein equations one can be solutions of
(\ref{ee2}) resulting in $\alpha _{2}=\beta =0.$ In such cases, $w_{2}$ can
be arbitrary functions on variables $(x^{2},v)$ with finite values for
derivatives in the limits $\alpha _{2},\beta \rightarrow 0$ eliminating the
''ill--defined'' situation $w_{2}\rightarrow 0/0.$ For the Ricci flow
equations with nonzero values of $\lambda _{\lbrack v]},$ such difficulties
do not arise. The second subset of N--connection (off--diagonal metric)
coefficients $n_{2}$ can be computed by integrating two times on variable $v$
in (\ref{ee4}), for given values $h_{3}$ and $h_{4}.$ One obtains%
\begin{equation}
n_{2}=n_{2[1]}(x^{2})+n_{2[2]}(x^{2})\ \hat{n}_{2}(x^{2},v),  \label{see4}
\end{equation}%
where
\begin{eqnarray*}
\hat{n}_{2}(x^{2},v) &=&\int h_{3}(\sqrt{|h_{4}|})^{-3}dv,  h_{4}^{\ast
}\neq 0;  \\
&=&\int h_{3}dv, h_{4}^{\ast }=0; \\
&=&\int (\sqrt{|h_{4}|})^{-3}dv, h_{3}^{\ast }=0,
\end{eqnarray*}%
and $n_{2[1]}(x^{2})$ and $n_{2[2]}(x^{2})$ are integration functions.

We conclude that any solution $\left( h_{3},h_{4}\right) $ of the equation (%
\ref{ee2}) with $h_{4}^{\ast }\neq 0$ and non--vanishing $\lambda _{\lbrack
v]} $ generates the solutions (\ref{see3}) and (\ref{see4}), respectively,
of equations (\ref{ee3}) and (\ref{ee4}). Such solutions (of the Einstein
equations) are defined by the mentioned classes of integration functions and
prescribed values for $b(x^{2},v)$ and $\psi (x^{2}).$ Further restrictions
on $(\epsilon ,g_{2})$ and $\left( h_{3},h_{4}\right) $ are necessary in
order to satisfy the equations (\ref{rfe1}) and (\ref{rfe2}) relating flows
of the metric and N--connection coefficients in a compatible manner. It is
not possible to solve in a quite general form such equations, but in the
next section we shall give certain examples of such solutions defining flows
of the Taub-NUT like metrics.

\subsection{Extracting solutions for the Levi-Civita connection}

The method outlined in the previous section allows us to construct integral
varieties for the Ricci flow equations (\ref{ee2})--(\ref{rfe2}) derived for
the canonical d--connection with nontrivial torsion, see formulas (56) and
(52) in Appendix to Ref. \cite{rf4de}. We can restrict such integral
varieties (constraining the off--diagonal metric, equivalently,
N--connection coefficients $w_{2}$ and $n_{2}$ and related integration
functions) in order to generate solutions for the Levi-Civita connection.
The conditions $_{\shortmid }\Gamma _{\beta \gamma }^{\alpha }=\widehat{%
\mathbf{\Gamma }}_{\ \alpha \beta }^{\gamma }$ (i.e. the coefficients of the
Levi-Civita connection are equal to the coefficients of the canonical
d--connection, both classes of coefficients being computed with respect to
the N--adapted bases (\ref{ddif}) and (\ref{dder})) hold true if there are
satisfied the equations (see (60) in Appendix to Ref. \cite{rf4de})\footnote{%
We emphasize that the connections on (pseudo) Riemannian and/or
Riemann--Cartan spaces are not defined as tensor objects. If their
coefficients are equal with respect to one frame, they can be very different
with respect to other frames.}:

\begin{eqnarray}
\frac{\partial h_{3}}{\partial x^{2}}-w_{2}h_{3}^{\ast }-2w_{2}^{\ast }h_{3}
&=&0,  \label{cond3a} \\
\frac{\partial h_{4}}{\partial x^{2}}-w_{2}h_{4}^{\ast } &=&0,
\label{cond3b} \\
n_{2}^{\ast }h_{4} &=&0.  \label{cond3c}
\end{eqnarray}%
The relations (\ref{cond3a}) and (\ref{cond3b}) are equivalent for the
general solutions $h_{3},$ see (\ref{solh5}), $h_{4},$ see (\ref{solh4}) and
$w_{2},$ see (\ref{see3}), generated by a function $\phi (x^{2},v)$ (\ref%
{afa1}) if $\phi \rightarrow \phi -\ln 2,$ when
\begin{equation*}
\phi =\ln |\left( \sqrt{|h_{4}|}\right) ^{\ast }|-\ln |\left( \sqrt{|h_{3}|}%
\right) |
\end{equation*}%
and
\begin{equation*}
w_{2}=(h_{4}^{\ast })^{-1}\frac{\partial h_{4}}{\partial x^{2}}=-(\phi
^{\ast })^{-1}\frac{\partial \phi }{\partial x^{2}},
\end{equation*}%
where $\phi =const$ is possible only for the vacuum Einstein solutions. In a
particular case, we can consider any parametrization of type $w_{2}=\widehat{%
w}_{2}(x^{2})q(v)$ for some functions $\widehat{w}_{k}(x^{i})$ and $q(v).$
The condition (\ref{cond3c}) for $h_{4}\neq 0$ constrains $n_{3}^{\ast }=0$
which holds true if we put the integration functions $n_{2[2]}=0$ in (\ref%
{see4}), when $n_{2}=n_{2[1]}(x^{2}).$

The final conclusion in this section is that taking any solution of
equations (\ref{ee2}), (\ref{ee3}) and (\ref{ee4}) we can restrict the
integral varieties to such integration functions satisfying the conditions
when the torsionless configurations for the Levi-Civita connection  are
extracted.

\section{Nonholonomic 3D Ricci Flows and Taub--NUT--dS Metrics}

We analyze the anholonomic frame method of constructing solutions defining
Ricci flows and exact solutions for three dimensional (3D) spacetimes with
negative cosmological constant $\lambda =-1/l^{2}.$ The primary metrics are
those for a $U(1)$ fibration over 2D spaces with constant curvature. By
nonholonomic deformations we shall transform such spaces into 3D manifolds,
or foliations (because in this case the nonholonomic structure is
integrable), with effective cosmological ''constant'' (anisotropically
depending on some coordinates, or running in time) polarized by string
corrections, Ricci flows, nontrivial torsion contributions, ....

\subsection{Solutions for 3D Ricci flows}

We deform a primary metric $\ \mathbf{\check{g}=}\left[ \check{g}_{2},\check{%
h}_{a},\check{N}_{2}^{a}\right] $ (in the next section such coefficients
will be stated to define certain 3D Taub-NUT like metrics) by considering
polarizations coefficients $\eta _{2},\eta _{a},\eta _{2}^{a}$ resulting in
the coefficients of ansatz (\ref{m2a}),
\begin{equation}
g_{2}=\eta _{2}(x^{2},v)\check{g}_{2},\ h_{a}=\eta _{a}(x^{2},v)\check{h}%
_{a},\ N_{2}^{a}=\eta _{2}^{a}(x^{2},v)\check{N}_{2}^{a}.  \label{polf3d}
\end{equation}%
In explicit form, such coefficients will define nonholonomic 3D Ricci flows
of certain type primary metrics. In this section we shall consider details
and examples on constructing nonholonomic Ricci flow solutions.

\subsubsection{General solutions for the Ricci flow equations}

The set of solutions of (\ref{ee2}) is parametrized by any functions $%
h_{3}(x^{2},v)$ and $h_{4}(x^{2},v)$ related by the condition (\ref{solh5}),
i.e. when
\begin{equation}
|h_{3}|=4e^{-2\phi (x^{2},v)}\left[ \left( \sqrt{|h_{4}|}\right) ^{\ast }%
\right] ^{2},  \label{s3dg2a}
\end{equation}%
for $h_{4}^{\ast }\neq 0$ and $\phi (x^{2},v)$ is a function to be computed
from
\begin{equation}
|h_{4}^{\ast }|=-(e^{\phi })^{\ast }/4\lambda _{\lbrack v]}  \label{s3dg2b}
\end{equation}%
where $\lambda _{\lbrack v]}(x^{2},v)$ is the ''vertically polarized
cosmological constant. If $\lambda _{\lbrack v]}\rightarrow 0,$ we have to
take $\phi \rightarrow 0$ such way that $h_{4}^{\ast }$ does not vanish.

The N--connection coefficients $w_{2}$ and $n_{2}$ are respectively defined
by formulas (\ref{see3}) and (\ref{see4}), when
\begin{equation}
w_{2}=-\partial _{2}\phi /\phi ^{\ast }  \label{s3dg3}
\end{equation}%
and
\begin{equation}
n_{2}=n_{2[1]}(x^{2})+n_{2[2]}(x^{2})\ \hat{n}_{2}(x^{2},v),  \label{s3dg4}
\end{equation}%
where
\begin{eqnarray*}
\hat{n}_{2}(x^{2},v) &=&\int h_{3}(\sqrt{|h_{4}|})^{-3}dv, h_{4}^{\ast
}\neq 0;  \\
&=&\int h_{3}dv, h_{4}^{\ast }=0; \\
&=&\int (\sqrt{|h_{4}|})^{-3}dv, h_{3}^{\ast }=0.
\end{eqnarray*}

One should be noted that the coefficients (\ref{s3dg2a}), (\ref{s3dg2b}), (%
\ref{s3dg3}) and (\ref{s3dg4}) for the ansatz (\ref{m2a}) were computed to
define exact solutions for the 3D Einstein equations with prescribed
polarizations of cosmological constants, for the canonical d--connection. We
can extract 3D (pseudo) Riemannian foliations if we impose further
constraints on the N--connection coefficients in order to extract
torsionless configurations for the Levi-Civita connection. This is possible
for any parametrizations of type $w_{2}=\check{w}_{2}(x^{2})q(v)$ and if the
integration function $n_{2[2]}(x^{2})$ is stated to be zero.

The flow equations for the 3D ansatz (\ref{m2a}), derived from the equations
(\ref{rfe1}) and (\ref{rfe2}),
\begin{eqnarray}
\frac{\partial }{\partial \tau }g_{2} &=&-2w_{2}h_{3}\frac{\partial }{%
\partial \tau }w_{2}-2n_{2}h_{4}\frac{\partial }{\partial \tau }n_{2},
\label{rfe13d} \\
\frac{\partial }{\partial \tau }h_{a} &=&2\lambda _{\lbrack v]}(x^{2},v)\
h_{a}.  \label{rfe23d}
\end{eqnarray}%
In explicit form, families of solutions of these equations can be generated
by fixing $\tau =x^{2},$ or $\tau =v,$ and integrating the equations for
certain prescribed values $\lambda _{\lbrack v]}(x^{2},v).$ In a particular
case, we can state that the target solutions define spacetimes with
effective cosmological constant induce from string gravity, when $\lambda
_{\lbrack h]}=\lambda _{\lbrack v]}=-\frac{\lambda _{\lbrack H]}^{2}}{4}$
(see Appendix in Ref. \cite{rf4de}); we have to solve the equations (\ref%
{rfe13d}) and (\ref{rfe23d}) for a such type prescribed cosmological
constant. Inversely, we can consider that $\lambda _{\lbrack v]}(x^{2},v)$
defines a nonhomogeneous polarization of the cosmological constant $\lambda
=-1/l^{2}$ in 3D gravity, defining self--consistent non trivial Ricci flows
under nonholonomic transforms.

We can chose the function $\phi (x^{2},v)$ from (\ref{s3dg2a}) and (\ref%
{s3dg2b}) to have $|h_{3}|=|h_{4}|.$\footnote{%
For 3D solutions, this can be achieved by a corresponding 2D coordinate
transform of $x^{2}$ and $v.$} For such parametrization, the equations (\ref%
{rfe13d}) and (\ref{rfe23d}) simplify substantially,
\begin{eqnarray*}
\frac{\partial }{\partial \tau } g_{2}(x^{2}) &=&-h_{3}\frac{\partial }{%
\partial \tau }\left\{ \left[ w_{2}(x^{2},v)\right] ^{2}\pm \left[
n_{2}(x^{2},v)\right] ^{2}\right\} , \\
\frac{\partial }{\partial \tau }\ln \left| \sqrt{|h_{3}(x^{2},v)|}\right|
&=&\lambda _{\lbrack v]}(x^{2},v),
\end{eqnarray*}%
where we take the sign ''+'' if the coordinates $y^{3}$ and $y^{4}$ have the
same signature, or ''$-"$ for different signatures. For $n_{2}=0,$ which is
always possible if we state that the integration functions in (\ref{s3dg4})
are zero (in this case we generate the Levi-Civita configurations), we can
define in explicit form certain classes of exact solutions derived for any
prescribed values of generating function $\phi (x^{2},v)$ and $\lambda
_{\lbrack h]}(x^{2}).$ The simplest approach is to compute the effective
vertical polarization of the cosmological constant, i.e. the function $%
\lambda _{\lbrack v]}(x^{2},v).$ Here it should be noted that the type of
solutions depends on the fact if the variable $\tau $ is holonomic or
nonholonomic.

\paragraph{3D Ricci flows on holonomic coordinate $x^{2}$:}

~

If the flow coordinate is taken to be the holonomic one, $\tau =x^{2},$ we
obtain
\begin{eqnarray}
-\frac{\partial }{\partial x^{2}}\ g_{2}(x^{2}) &=&h_{3}(x^{2},v)\frac{%
\partial }{\partial x^{2}}\left\{ \left[ w_{2}(x^{2},v)\right] ^{2}\right\} ,
\notag \\
\frac{\partial }{\partial x^{2}}\ln \left| \sqrt{|h_{3}(x^{2},v)|}\right|
&=&\lambda _{\lbrack v]}(x^{2},v).  \label{rf3dh}
\end{eqnarray}%
We search a class of solutions of equations (\ref{rf3dh}) with separation of
variables in the form%
\begin{equation}
h_{3}=h_{4}=A(x^{2})B(v)  \label{apaux3}
\end{equation}%
when
\begin{equation*}
\check{\phi}=\phi -\ln 2=\ln \left| \partial _{v}\ln \sqrt{|h_{3}|}\right|
\end{equation*}%
and $w_{2}=-\partial _{2}\check{\phi}/\partial _{v}\check{\phi}.$ We obtain
from the first equation that%
\begin{equation}
|A|=A_{0}+\frac{1}{8c_{0}}\int \sqrt{|\partial _{2}g_{2}(x^{2})|}dx^{2}
\label{apaux1}
\end{equation}%
and $B$ must solve the equation%
\begin{equation*}
BB^{\ast \ast }+c_{0}(B^{\ast })^{3}-B^{2}(B^{\ast })^{2}=0
\end{equation*}%
for some integration constants $c_{0}$ and $A_{0}.$ For $B^{\ast }=\partial
_{v}B\neq 0,$ we write the last equation as%
\begin{equation*}
\left( \ln |B^{\ast }|\right) ^{\ast }+c_{0}B^{\ast }(\ln |B|)^{\ast }=(\ln
|B|)^{\ast }.
\end{equation*}%
Introducing the function $H=(\ln |B|)^{\ast },$ the equation transform into
\begin{equation*}
\frac{H^{\ast }}{H^{2}}+c_{0}B^{\ast }=0
\end{equation*}%
which can be integrated on $v,$ and than transformed into
\begin{equation*}
B^{\ast }=\frac{B}{c_{1}-c_{0}B}
\end{equation*}%
with the solution
\begin{equation}
c_{1}\ln |B(v)|-c_{0}B(v)=v+v_{0},  \label{apaux2}
\end{equation}%
for some integration constants $c_{0},c_{1}$ and $v_{0.}$

Introducing $h_{3}=AB$ (\ref{apaux3}), for $A$ defined by (\ref{apaux1}) and
$B$ defined by (\ref{apaux2}), into the second equation (\ref{rf3dh}), we
get that such solutions can be constructed for any polarized on $x^{2}$
vertical cosmological constant
\begin{equation*}
\lambda _{\lbrack v]}(x^{2})=\partial _{2}\ln \sqrt{|A|}.
\end{equation*}

We conclude that 3D Ricci flow solutions on holonomic variable $x^{2},$ with
separation of variables can be generated for any polarized anisotropically
cosmological constants with any $\lambda _{\lbrack v]}(x^{2}).$ As a matter
of principle we can consider dependencies of type $\lambda _{\lbrack
v]}(x^{2},v)$ for certain configurations not admitting separation of
variables.

Putting together the above formulas, we define a class of 3D metrics solving
the system (\ref{rf3dh}),
\begin{equation}
\ \mathbf{g}=g_{2}(x^{2})(dx^{2})^{2}+A(x^{2})B(v)\left[ \epsilon
_{3}(dv+w_{2}(x^{2},v)\ dx^{2})^{2}+\epsilon _{4}(dy^{4})^{2}\right]
\label{srf3dh}
\end{equation}%
where $g_{2}(x^{2})$ is an arbitrary function. There are considered certain
integration constants $c_{0},c_{1}$ and $A_{0}$ to be defined by fixing a 3D
local coordinate system and certain boundary conditions. The nontrivial
N--connection coefficient is given by
\begin{equation*}
w_{2}(x^{2},v)=-\partial _{2}\ln |A(x^{2})|/\partial _{v}\ln |B(v)|.
\end{equation*}

The metric (\ref{srf3dh}) describes families of Ricci flow solutions for any
prescribed values of $g_{2}(x^{2})$ and any polarizations of cosmological
constant\newline
$\lambda _{\lbrack v]}(x^{2})=\partial _{2}\ln \sqrt{|A(x^{2})|}.$ The
physical meaning of such solutions is that they describe Ricci flows on
coordinate $x^{2}$ of metrics of type (\ref{m2a}) with the Levi-Civita
connection (when $n_{2}=0$) as flows of 3D Einstein spacetimes with
effective polarizations of the cosmological constant. In a particular case,
we can say that $\lambda _{\lbrack v]}(x^{2})$ is defined by any oscillating
function, or one dimensional solitonic waves with a time like coordinate $%
x^{2}.$ Such flows are with self--consistent (preserved during the flow
evolution) separation of variables.

\paragraph{3D Ricci flows on nonholonomic coordinate $v$:}

~

For $\tau =v,$ the Ricci flows are directed by dependencies on the
anholonomic coordinate,
\begin{eqnarray}
0 &=&\frac{\partial }{\partial v}\left\{ \left[ w_{2}(x^{2},v)\right]
^{2}\pm \left[ n_{2}(x^{2},v)\right] ^{2}\right\} ,  \label{rf3dah} \\
\frac{\partial }{\partial v}\ln \left| \sqrt{|h_{3}(x^{2},v)|}\right|
&=&\lambda _{\lbrack v]}(x^{2},v),  \notag
\end{eqnarray}%
where $\partial _{v}n_{2}=0,$ because $n_{2}=n_{2[1]}(x^{2}),$ for the Ricci
flows of Levi-Civita configurations.

The solution with separation of variables of system (\ref{rf3dah}) when $%
h_{3}=A_{1}(x^{2})B(v)$ can be performed similarly as for (\ref{rf3dh}) but
with $|A_{1}|=A_{0}=const$ introduced into $B(v)$ being determined by the
same type of solution like (\ref{apaux2}). The second difference from the
previous case is that there are admitted solutions for vertically polarized
on $v$ cosmological constants when%
\begin{equation*}
\lambda _{\lbrack v]}(v)=\partial _{v}\ln \sqrt{|B(v)|}.
\end{equation*}%
This allows us to conclude that nonholonomic 3D Ricci flows with separation
variables are possible for vertical polarizations of the cosmological
constant, i.e. for any $\lambda _{\lbrack v]}(v).$ In this case, we can also
consider dependencies of type $\lambda _{\lbrack v]}(x^{2},v)$ for certain
configurations not admitting separation of variables.

Finally, we summarize that the solutions of (\ref{rf3dah}) are parametrized
\begin{equation}
\ \mathbf{g}=g_{2}(x^{2})(dx^{2})^{2}+B(v)\left[ \epsilon
_{3}dv^{2}+\epsilon _{4}(dy^{4})^{2}\right] ,  \label{srf3dah}
\end{equation}%
when $|h_{3}|=B(v)$ (\ref{apaux2}). For this metric, $w_{2}(x^{2},v)=0$ and%
\newline
$\lambda _{\lbrack v]}(v)=\partial _{v}\ln \sqrt{|B(v)|}.$ The Ricci flows
described by the metric (\ref{srf3dah}) are on nonholonomic coordinate $v$
for any $\lambda _{\lbrack v]}(v).$ In such cases, the cosmological constant
anisotropically runs on nonholonomic variable $v$ in the vertical 2D
subspace.

\subsubsection{3D Ricci flows of Taub Nut metrics with holonomic time like
coordinate}

We consider the primary ansatz
\begin{equation}
d\check{s}^{2}=\epsilon _{2}\check{g}_{2}(x^{2},v)(dx^{2})^{2}+\epsilon _{3}%
\check{h}_{3}(dv)^{2}+\epsilon _{4}\check{h}_{4}\left[ dy^{4}+\check{n}%
_{2}(x^{2})dx^{2}\right] ^{2}.  \label{3dhcansp}
\end{equation}%
For the data
\begin{eqnarray*}
x^{2} &=&t,y^{3}=r,y^{4}=\vartheta ; \epsilon _{2}=-1,\epsilon
_{3}=\epsilon _{4}=1; \\
\check{h}_{3} &=&l^{2}/4, \check{h}_{4}=l^{2}/16n^{2};  \\
\check{g}_{2} &=&\left\{
\begin{array}{c}
(l^{2}/4) \sinh ^{2}r; \\
(l^{2}/4) \cosh ^{2}r; \\
(l^{2}/4) e^{2r};%
\end{array}%
\right. \mbox{\ and\  } \check{n}_{2}=\left\{
\begin{array}{c}
-2n \cosh r; \\
-2n \sinh r; \\
-2n e^{r};%
\end{array}%
\right.
\end{eqnarray*}%
we get three classes of exact solutions of 3D Einstein field equations with
negative cosmological constant $\lambda =-1/l^{2}$ and nut parameter $n$
considered in Ref. \cite{ms} and corresponding to some Lorentzian versions
of the so--called Thurston's geometries \cite{thurs}. We analyze nonhlonomic
Ricci flows of such geometries, on holonomic coordinate $x^{2}=t.$

Applying a conformal map of (\ref{3dhcansp})
\begin{equation*}
d\check{s}^{2}\rightarrow d\check{s}_{(c)}^{2}=\left( \check{g}_{2}\right)
^{-1} d\check{s}^{2}
\end{equation*}%
and then a nonholonomic transform (the conformal transform is necessary in
order to define the nonholonomic deformations in a simplified form), one
generates the target ansatz
\begin{eqnarray}
ds^{2} &=&\epsilon _{2}\eta _{2}(t,r)(dt)^{2}+\eta _{3}(t,r)\frac{\check{h}%
_{3}}{\check{g}_{2}(r)}dr^{2}  \label{3dhcanspd} \\
&&+\eta _{4}(t,r)\frac{\check{h}_{4}}{\check{g}_{2}(r)}\left[ d\vartheta
+\eta _{2}^{4}(t,r)\ \check{n}_{2}(r)\ dt\right] ^{2}  \notag
\end{eqnarray}%
which is of type (\ref{m2a}) with the coefficients considered to be certain
functions (\ref{s3dg2a}), (\ref{s3dg2b}), (\ref{s3dg3}) and (\ref{s3dg4})
induced by nontrivial polarizations $\eta _{2}(t,r),\eta _{3}(t,r),$ $\eta
_{2}^{4}(t,r)$ and $w_{2}(t,r)$ in order to solve the 3D equations (\ref{ee2}%
)--(\ref{ee4}) and the flow equation (\ref{rf3dh}).

For a restricted class of polarizations when $\eta _{2}^{4}=0,\eta _{2}=\eta
_{2}(t)$ and
\begin{equation*}
g_{2}(t)=-\eta _{2}(t),h_{3}=\eta _{3}(t,r)\frac{\check{h}_{3}}{\check{g}%
_{2}(r)}=\eta _{4}(t,r)\frac{\check{h}_{4}}{\check{g}_{2}(r)},
\end{equation*}%
we can use the ansatz of type (\ref{srf3dah})
\begin{equation}
\ \mathbf{g}=g_{2}(t)(dt)^{2}+A(t)B(r)dr^{2}+(d\vartheta )^{2}
\label{sol3dth}
\end{equation}%
where $|h_{3}|=A(t)B(r),$
\begin{equation*}
A=A_{0}+\frac{1}{8c_{0}}\int \sqrt{|\partial _{t}g_{2}(t)|}dt
\end{equation*}%
and $B$ is defined from
\begin{equation*}
c_{1}\ln |B(r)|-c_{0}B(r)=r+r_{0},
\end{equation*}%
see formula (\ref{apaux2}). The nontrivial N--connection coefficient in (\ref%
{sol3dth}) is computed as
\begin{equation*}
w_{2}(t,r)=-\partial _{t}\ln |A(t)|/\partial _{r}\ln |B(r)|.
\end{equation*}%
The integration constants in the above formulas are denoted $%
c_{0},c_{1},A_{0},r_{0},g_{2(0)}$ and $\psi _{0}.$

The metric (\ref{sol3dth}) describes families of Ricci flows of the 3D
Taub-NUT like solutions for any prescribed values $g_{2}(t)$ and
polarization of
cosmological constant $\lambda _{\lbrack v]}(t)=\partial _{t}\ln \sqrt{|A(t)|%
}.$ We can consider more particular cases when $\lambda _{\lbrack v]}(t)$ is
any oscillating function, or one dimensional solitonic wave on the time like
coordinate $t.$ Such flows preserve the separation of variables and the
vanishing torsion for the Levi-Civita connection.

\subsubsection{3D Ricci flows on nonholonomic time like coordinate}

We consider another class of Ricci flows of 3D metrics. We take the primary
ansatz in the form (\ref{3dhcansp}) but with a reparametrization of
coordinates $(t\rightarrow i\chi ,\vartheta \rightarrow i\vartheta
,r\rightarrow it)$ and different signature and parametrization of the
nontrivial metric and N--connection coefficients,%
\begin{eqnarray}
x^{3} &=&\vartheta ,y^{3}=v=t,y^{4}=\chi ;\ \epsilon _{2}=-1,\epsilon
_{3}=1,\epsilon _{4}=1;  \label{datas3a} \\
\check{g}_{2}(t) &=&\frac{l^{2}}{4}\sin ^{2}t,\ \check{h}_{3}=\frac{l^{2}}{4}%
,\ \check{h}_{4}=\frac{l^{2}}{16n^{2}},\ \check{n}_{2}(t)=2n\cos t.  \notag
\end{eqnarray}%
The quadratic element $d\check{s}^{2}$ defined by the data define another
class of Thors\-ton's geometries and exact solutions of 3D Einstein
equations with negative cosmological constant also considered in Ref. \cite%
{ms}. Transforming conformally the primary metric, $d\check{s}%
^{2}\rightarrow d\check{s}_{(c)}^{2}=\left( \check{g}_{2}\right) ^{-1}\ d%
\check{s}^{2},$ and then applying a nonholonomic transform, one generates
the target ansatz
\begin{eqnarray}
ds^{2} &=&\epsilon _{2}\eta _{2}(\vartheta )(d\vartheta )^{2}+\epsilon
_{3}\eta _{3}(\vartheta ,t)\frac{\check{h}_{3}}{\check{g}_{2}(t)}dt^{2}
\notag \\
&&+\epsilon _{4}\eta _{4}(\vartheta ,t)\frac{\check{h}_{4}}{\check{g}_{2}(t)}%
\left[ d\chi +\eta _{2}^{4}(\vartheta ,t)\ \check{n}_{2}(t)d\vartheta \right]
^{2}  \label{3dhcanspd2}
\end{eqnarray}%
which is of type (\ref{m2a}) with the coefficients considered to be certain
functions (\ref{s3dg2a}), (\ref{s3dg2b}), (\ref{s3dg3}) and (\ref{s3dg4})
induced by nontrivial polarizations $\eta _{2}(\vartheta ),\eta
_{3}(\vartheta ,t),$ $\eta _{2}^{4}(\vartheta ,t)$ and $w_{2}(\vartheta ,t)$
in order to solve the 3D equations (\ref{ee2})--(\ref{ee4}) and the flow
equation (\ref{rf3dah}). This ansatz also may define Ricci flows on time
like coordinate $t$ but in a very different form than (\ref{3dhcanspd}): In
this case we shall have an angular anisotropy on $\vartheta $ and locally
anisotropic flows on time $t$ (in the previous example the flow coordinate
was holonomic for the equation (\ref{rf3dh})).

The method of constructing the solutions of (\ref{ee2})--(\ref{ee4}) for the
ansatz (\ref{3dhcanspd2}) is completely similar to that considered in the
previous example. For simplicity, we shall omit details and write down the
solution applying formulas (\ref{s3dg2a}), (\ref{s3dg2b}), (\ref{s3dg3})\
and (\ref{s3dg4}) stated for the data (\ref{datas3a}) and
\begin{eqnarray}
g_{2} &=&\epsilon _{2}\eta _{2}(\vartheta ),\ h_{3}=\epsilon _{3}\eta
_{3}(\vartheta ,t)\frac{\check{h}_{3}}{\check{g}_{2}(t)},\ h_{4}=\epsilon
_{4}\eta _{4}(\vartheta ,t)\frac{\check{h}_{4}}{\check{g}_{2}(t)}  \notag \\
N_{2}^{3} &=&0,\ N_{2}^{4}=n_{2}(\vartheta ,t)=\eta _{2}^{4}(\vartheta ,t)\
\check{n}_{2}(t),  \label{pol3dex2}
\end{eqnarray}%
when $x^{2}=\vartheta $ and $y^{3}=v=t.$ The solutions for Ricci flows with
separation of variables are of type (\ref{srf3dah}) parametrized in the form
\begin{equation}
\ \mathbf{g}=g_{2}(\vartheta )(d\vartheta )^{2}+B(t)\left[ dt^{2}+(d\chi
)^{2}\right] ,  \label{sol3dtah}
\end{equation}%
when $h_{3}=h_{4}=B(t)$ with $B(t)$ defined from%
\begin{equation*}
c_{1}\ln |B(t)|-c_{0}B(t)=t+t_{0},
\end{equation*}%
see formula (\ref{apaux2}). For this metric $w_{2}(\vartheta ,t)=0$ and $%
\lambda _{\lbrack v]}(t)=\partial _{t}\ln \sqrt{|B(t)|}.$

The Ricci flows defined by the solutions (\ref{sol3dtah}) are on
nonholonomic time coordinate $t.$ They are defined for any coefficient $%
g_{2}(\vartheta )$ and any $\lambda _{\lbrack v]}(t)$ running in time in a
manner compatible with $h_{3}=h_{4}.$ For such solutions, the cosmological
constant is anisotropically polarized on angular coordinate $\vartheta $ in
the ''horizontal'' direction and runs on nonholonomic time variable $t$ in
the vertical 2D subspace. The class of metrics (\ref{sol3dtah}) describes
Ricci flows of conformally deformed Thorston's geometries stated by
polarizations (\ref{pol3dex2}) and data (\ref{datas3a}).

\subsection{New classes of 3D exact solutions of Einstein equations}

There is a subclass of Thorston's geometries parametrized by stationary
metrics which under nonholonomic deformations transform into other classes
of exact solutions defining nonholonomic (foliated) 3D spacetimes. In this
case we consider only solutions of the equations (\ref{ee2})--(\ref{ee4})
but do not subject the metric and N--connection coefficients to solve the
equations (\ref{rfe1}) and (\ref{rfe2}).

We consider the primary ansatz
\begin{equation}
d\check{s}^{2}=\epsilon _{2}\check{g}_{2}(v)(dx^{2})^{2}+\epsilon _{3}\check{%
h}_{3}(dv)^{2}+\epsilon _{4}\check{h}_{4}\left[ dy^{4}+\check{n}%
_{2}(x^{2})dx^{2}\right] ^{2}.  \label{pm3}
\end{equation}%
For the data
\begin{eqnarray*}
x^{2} &=&\vartheta ,y^{3}=r,y^{4}=t; \epsilon _{2}=1,\epsilon
_{3}=1,\epsilon _{4}=-1; \\
\check{h}_{3} &=&l^{2}/4, \check{h}_{4}=l^{2}/16n^{2};  \\
\check{g}_{2} &=&\left\{
\begin{array}{c}
(l^{2}/4) \sinh ^{2}r; \\
(l^{2}/4) \cosh ^{2}r; \\
(l^{2}/4) e^{2r};%
\end{array}%
\right. \mbox{\ and\  } \check{n}_{2}=\left\{
\begin{array}{c}
-2n \cosh r; \\
-2n \sinh r; \\
-2n e^{r};%
\end{array}%
\right.
\end{eqnarray*}%
we get other three classes of exact solutions of 3D Einstein field equations
with negative cosmological constant $\lambda =-1/l^{2}$ and nut parameter $n$
considered in Ref. \cite{ms} and corresponding to some Lorentzian versions
of the so--called Thurston's geometries \cite{thurs} (we already considered
different types of primary 3D exact solutions given by ansatz (\ref{3dhcansp}%
) and data (\ref{datas3a})). Then we introduce a conformal map $d\check{s}%
^{2}\rightarrow d\check{s}_{(c)}^{2}=\left[ \check{g}_{2}(v)\right] ^{-1}d%
\check{s}^{2}$ and then a nonholonomic deformation to the off--diagonal
ansatz%
\begin{eqnarray*}
ds^{2} &=&(dx^{1})^{2}+\eta _{2}(\vartheta )(d\vartheta )^{2}+\eta
_{3}(\vartheta ,r)\frac{\check{h}_{3}}{\check{g}_{2}(r)}\left[
dr+w_{2}(\vartheta ,r)d\vartheta \right] ^{2} \\
&&-\eta _{4}(\vartheta ,r)\frac{\check{h}_{4}}{\check{g}_{2}(r)}\left[
dt+\eta _{3}^{4}(\vartheta ,t)\check{n}_{2}(r)d\vartheta \right] ^{2}
\end{eqnarray*}%
were we trivially embedded the 3D ansatz into a 4D (this is necessary in
order to consider cosmological constants induced from string gravity). This
metric is of type (\ref{m2a}) with polarization functions (\ref{polf3d}) and
N--connection coefficients parametrized in the form
\begin{eqnarray}
g_{1} &=&1,\ g_{2}=\eta _{2}(\vartheta ),\ h_{3}=\eta _{3}(\vartheta ,r)%
\frac{\check{h}_{3}}{\check{g}_{2}(r)},\ h_{4}=\eta _{4}(\vartheta ,r)\frac{%
\check{h}_{4}}{\check{g}_{2}(r)},  \label{pol3des} \\
\ N_{2}^{3} &=&w_{2}(\vartheta ,r),\ N_{2}^{4}=n_{2}(\vartheta ,r)=\eta
_{2}^{4}(\vartheta ,r)\check{n}_{2}(r).  \notag
\end{eqnarray}%
We can use formulas (\ref{s3dg2a}), (\ref{s3dg2b}), (\ref{s3dg3}) and (\ref%
{s3dg4}) in order to write down the general solution of the Einstein
equations with effective cosmological constant induced from string gravity.
The solutions for $\lambda _{\lbrack v]}=-\lambda _{H}^{2}/4$ are given by
the coefficients of the d--metric
\begin{eqnarray}
g_{1} &=&1,\ g_{2}(\vartheta )=\eta _{2}(\vartheta ),  \label{data3dm} \\
|h_{3}| &=&\left[ \partial _{r}\phi (\vartheta ,r)\right] ^{2}e^{-\phi
(\vartheta ,r)}\frac{\check{h}_{4}}{\lambda _{H}^{2}\check{g}_{2}(r)},\
h_{4}=\frac{\varepsilon _{4}}{\lambda _{H}^{2}}e^{\phi (\vartheta ,r)}\frac{%
\check{h}_{4}}{\check{g}_{2}(r)},  \notag
\end{eqnarray}%
where $\psi _{0}$ and $g_{2(0)}$ are integration constants and $\phi
(\vartheta ,r)$ is an arbitrary function, and of N--connection coefficients
\begin{equation*}
w_{2}=-\partial _{\vartheta }\phi (\vartheta ,r)/\partial _{r}\phi
(\vartheta ,r)
\end{equation*}%
and
\begin{equation}
n_{2}=n_{2[1]}(\vartheta )+n_{2[2]}(\vartheta )\ \hat{n}_{k}(\vartheta ,r),
\label{data3dn}
\end{equation}%
where
\begin{equation*}
\hat{n}_{2}(\vartheta ,r)=\int h_{3}(\sqrt{|h_{4}|})^{-3}dr,\
\end{equation*}%
for $ h_{4}^{\ast },h_{3}^{\ast }\neq 0.$ We can define the polarization
functions by introducing (\ref{data3dm}) and (\ref{data3dn}) into (\ref%
{pol3des}).

Putting together the defined coefficients, we construct this class of 3D
exact solutions generated effectively in string gravity:%
\begin{eqnarray}
ds^{2} &=&(dx^{1})^{2}+g_{2}\ (d\vartheta )^{2}+\left( \partial _{r}\phi
\right) ^{2}e^{-\phi }\frac{\check{h}_{4}}{\lambda _{H}^{2}}\left[ dr-\frac{%
\partial _{\vartheta }\phi }{\partial _{r}\phi }d\vartheta \right] ^{2}
\notag \\
&&-e^{\phi }\frac{\check{h}_{4}}{\lambda _{H}^{2}\check{g}_{2}}\left[
dt+\left( n_{2[1]}+n_{2[2]}\ \int h_{3}(\sqrt{|h_{4}|})^{-3}dr\right)
\check{n}_{2}\ d\vartheta \right] ^{2}.  \label{sol3des}
\end{eqnarray}%
Such solutions are induced as nonholonomic string deformations of
conformally deformed Thurston's geometries stated by the primary metrics (%
\ref{pm3}). They define nonholonomic fibrations over 3D spacetime trivially
embedded into the 4D spacetime. This family of solutions is generated by
arbitrary integration functions $\phi (\vartheta ,r),n_{2[1]}(\vartheta )$
and $n_{2[2]}(\vartheta )$ and integration constants $\psi _{0}$ and $%
g_{2(0)}.$ The subclass of solutions with trivial torsion for the Levi
Civita connection (nevertheless with nontrivial string torsion) can be
extracted if we impose the condition $n_{2[2]}=0$ and chose, for instance, $%
\phi (\vartheta ,r)=\phi _{1}(\vartheta )\phi _{2}(r)$ to induce a
parametrization of type $w_{2}=\widehat{w}_{2}(\vartheta )q(r).$  This
defines certain spacetimes as 3D foliation structures. In the limit of
trivial polarizations $\eta \rightarrow 1$ and $w_{2}\rightarrow 0,$ with $%
\lambda _{\lbrack v]}=-\lambda _{H}^{2}/4\rightarrow -1/l^{2}$ the metric (%
\ref{sol3des}) does not transform into a solution of the 3D Einstein
equations with cosmological constant $\lambda =-1/l^{2}$ $\ $but into a
conformal transform of a such solution stated above by $d\check{s}_{(c)}^{2}.
$ This is because, as a matter of principle, the anholonomic frame method
can be applied to deform primary metrics with are not exact solutions.
Nevertheless, the final result is always related to certain classes of
generic off--diagonal solutions.

It should be noted that for 3D curved spaces any metric can be diagonalized
by corresponding coordinate transforms. This holds true for the generated
classes of solutions. We can not apply any type of coordinate transform if
we wont to preserve a prescribed nonholonomic/ foliated structure for new
classes of locally anisotropic Taub-NUT spacetimes. Finally, we note that
the solutions (\ref{sol3des}) can not be deformed (following the anhlonomic
frame method applied in this work) into certain solutions of the Ricci flow
equations because the time like coordinate for the considered family of
ansatz was chosen to be $y^{4}=t,$ when the primary and target metrics do
not depend on this variable by definition. So, such flow solutions can not
be constructed but stationary generic off-diagonal solutions of the Einstein
equations are possible.

\section{Outlook and Discussion}

We have considered here three dimensional solutions of the Ricci flow
equations, in a special case, defining flows of the Thorston's geometries
and corresponding Taub-NUT like metrics. These solutions were constructed
following the anholonomic frame method  and the geometry of 3D foliated
manifolds, in general, with  nontrivial torsion. The novelty of the method
is that it allows  to consider off--diagonal Ricci flows with possible
constrains  and nonholonomic deformations resulting in effectively
nonholomogeneous cosmological constants with anisotropic polarizations
modeling flows of the Einstein spaces. The solutions can be  generalized
for four dimensions as it is considered in the  partner paper \cite{rf4de}.

It was found that depending of the fact if the time like flow  coordinate is
holonomic or anholonomic the type of flows and  admissible polarizations of
the cosmological constants are very  different. In the first case there are
possible both, for  instance, angular and time dependencies of the
cosmological  constant but in the second case there are self--consistent
only  configurations with running in time cosmological "constants". For
certain classes of Thurston's geometries it is not possible to  generate
anholonomic Ricci flows, following our methods of  solutions. Nevertheless,
generalizations to new classes  of exact 3D solutions defining generic
off--diagonal Einstein  spaces can be obtained. Here, one should be noted
that even in 3D  every metric can be diagonalized by coordinate transforms,
the  off--diagonal metric terms have a special physical importance if
certain nonholonomic constraints and contributions of torsion,  for
instance, from string gravity are taken into consideration.

The method elaborated in this work can be applied for any  signatures of
metric and for various primary metrics and linear  connections (even they do
not define an exact solution)  nonholonomically deformed in order to
generate exact solutions of  the Ricci flow equations. In a number of cases,
there are  nontrivial limits of the flow solutions to certain classes of
exact solutions of the field equations.

We leave for future work the study of thermodynamic properties of  such
solutions which can be considered as some equilibrium states  of a
corresponding locally anisotropic space time kinetic model \cite{vap2,vsgg}.
It is also worth mentioning can be used as test  "flow" grounds for AdS/CFT
correspondence and various models with  nontrivial topology \cite{cejm,amo}.

\vskip5pt

\textbf{Acknowledgement: } S. V. is grateful to D. Singleton, E. Gaburov and
D. Gon\c ta for former collaboration and support.
 He thanks the Fields Institute for accepting his visit.
M. V. has been supported in part by the MEC-CEEX Program, Romania.


\begin{thebibliography}{99}
\bibitem{ham1} R. S. Hamilton, {\em  J. Diff. Geom.} \textbf{17} (1982) 255

\bibitem{ham2} R. S. Hamilton, in: {\it Surveys in Differential Geometry},
Vol. 2 (International Press, 1995), pp. 7--136

\bibitem{aubin} T. Aubin, {\it Some Nonlinear Problems in Riemannian
Geometry}, (Sprin\-ger Verlag, 1998)

\bibitem{cao} H.-D. Cao, B. Chow, S.-C.Chu and S.-T.Yau (Eds.), {\it Collected
Papers on Ricci Flow}, (International Press, Somerville, 2003)

\bibitem{chen} B.--L. Chen and K.--P. Zhu, {\it Uniqueness of the Ricci Flow on
Complete Noncompac Manifolds}, math.DG/ 0505447

\bibitem{per} G. Perelman, {\it The Entropy Formula for the Ricci Flow and its
Geometric Applications}, math.DG/ 0211159

\bibitem{bakas} I. Bakas, {\em JHEP} \textbf{0308} (2003) 013

\bibitem{geg} J. Gegenberg and G. Kunstatter, {\em Class. Quant. Grav.}
\textbf{21} (2004) 1197

\bibitem{dm} X. Dai, Li Ma, {\it Mass under the Ricci Flow}, math.DG/ 0510083

\bibitem{wg} W. Graf, {\it Ricci Flow Gravity}, gr--qc/ 0602054

\bibitem{hw} M. Headrick, T. Wiseman, {\em Int. J. Mod. Phys.  A} \textbf{22} (2007) 1135

\bibitem{rf4de} S. Vacaru and  M. Visinescu,
{\em Int. J. Mod. Phys.  A} \textbf{22} (2007) 1135


\bibitem{crvis} S. A. Carstea and M. Visinescu, {\em Mod. Phys. Lett. A} \textbf{20} (2005) 2993

\bibitem{vrf} S. Vacaru, {\em Int. J. Mod. Phys.  A} \textbf{21} (2006) 4899

\bibitem{thurs} J. Gegenberg, S. Vaidya and J. F. Vazquez--Poritz, {\em Class.
Quant. Grav.} \textbf{19 }(2002) L199

\bibitem{ms} R. Mann and C. Stelea, {\em Class. Quant. Grav.} \textbf{21}
(2004) 2937

\bibitem{vhep2} S. Vacaru, {\em JHEP} \textbf{04} (2001) 009

\bibitem{vs} S. Vacaru and D. Singleton, {\em J. Math. Phys.} \textbf{43} (2002)
2486

\bibitem{vesnc} S. Vacaru, {\em J. Math. Phys.} \textbf{46} (2005) 042503

\bibitem{vt} S. Vacaru and O. Tintareanu-Mircea, {\em Nucl. Phys. B}
\textbf{626} (2002) 239

\bibitem{vp} S. Vacaru and F. C. Popa, {\em Class. Quant. Grav.} \textbf{18}
(2001) 4921

\bibitem{vap2} S. Vacaru, {\em Ann. Phys. (N.Y.)} \textbf{290 } (2001) 83

\bibitem{vsgg} {\it Clifford and Riemann- Finsler Structures in Geometric
Mechanics and Gravity, Selected Works}, by S. Vacaru, P. Stavrinos, E.
Gaburov and D. Gon\c ta. Differential Geometry -- Dynamical Systems,
Monograph 7 (Geometry Balkan Press, 2006);
www.mathem.pub.ro/dgds/mono/va-t.pdf and gr-qc/0508023

\bibitem{cejm} A. Chamblin, R. Emparan, C. V. Johnston and R. C. Mayers,
{\em Phys. Rev.} \textbf{D59} (1999) 064010

\bibitem{amo} N. Alonco--Alberca, P. Meesen and T. Ortin, {\em Class. Quant.
Grav.} \textbf{17} (2000) 2783
\end{thebibliography}
\end{document}